\documentclass{INTERSPEECH2023}
\usepackage{amsmath,graphicx}
\usepackage{multirow,array,enumitem,adjustbox}
\usepackage{booktabs,threeparttable,makecell}
\usepackage{xcolor}
\interspeechcameraready

\title{Build a SRE Challenge System: Lessons from VoxSRC 2022 and CNSRC 2022}

\name{Zhengyang Chen$^{1,*}$, Bing Han$^{1,*}$, Xu Xiang$^{2}$, Houjun Huang$^{2}$, Bei Liu$^1$, Yanmin Qian$^{1,\dagger}$ \thanks{$^*$Equal Contribution} \thanks{$^\dagger$Corresponding Author} \thanks{This work was supported in part by China NSFC projects under Grants 62122050 and 62071288, and in part by Shanghai Municipal Science and Technology Major Project under Grant 2021SHZDZX0102.
}}
\address{
  $^1$MoE Key Lab of Artificial Intelligence, AI Institute \\
  X-LANCE Lab, Department of Computer Science and Engineering, Shanghai Jiao Tong University \\
  $^2$AISpeech Ltd, Suzhou, China}
\email{\{zhengyang.chen, hanbing97, yanminqian\}@sjtu.edu.cn}

\begin{document}

\maketitle
 
\begin{abstract}
Many speaker recognition challenges have been held to assess the speaker verification system in the wild and probe the performance limit. 
Voxceleb Speaker Recognition Challenge (VoxSRC), based on the voxceleb, is the most popular. Besides, another challenge called CN-Celeb Speaker Recognition Challenge (CNSRC) is also held this year, which is based on the Chinese celebrity multi-genre dataset CN-Celeb. 
Last year, our team participated in both speaker verification closed tracks in CNSRC 2022 and VoxSRC 2022, and achieved the 1$^{st}$ place and 3$^{rd}$ place respectively. 
In most system reports, the authors usually only provide a description of their systems but lack an effective analysis of their methods. 
In this paper, we will outline how to build a strong speaker verification challenge system and give a detailed analysis of each method compared with some other popular technical means.
\end{abstract}

\section{Introduction}
%  Speaker verification aims to verify one's identity based on his or her voice. Due to the boom of deep learning, the deep neural network based speaker verification method has completely replaced the traditional statistical methods \cite{} and has become the main field of study. Researchers have explored different neural networks \cite{}  to extract the speaker information from speech and explored different pooling functions \cite{} to map the variable-length audio features to fixed-dimension speaker embedding representation. Besides, a lot of attention has also been put on the loss function designing \cite{}, which aims to help the neural network extract robust and discriminative speaker embedding.

In order to evaluate how well current speaker recognition technology is able to identify speakers in unconstrained or ‘in the wild’ data and explore the performance limitation of the speaker recognition system, many speaker recognition challenges have been held in recent years~\cite{sadjadi20222021,brown2022voxsrc,qin2020interspeech}. 
Among these challenges, the most prestigious one is the Voxceleb Speaker Recognition Challenge (VoxSRC) \cite{brown2022voxsrc}, which is held once a year since 2019 and based on the very popular public speaker recognition dataset Voxceleb \cite{nagrani2017voxceleb,chung2018voxceleb2}. 
All the speech segments in the Voxceleb dataset are `real world' utterances that are collected from YouTube for several thousand individuals. 
Similar to Voxceleb, there is another dataset called CN-Celeb \cite{fan2020cn,li2022cn}, which is downloaded from Chinese social media websites. And the data collectors of CN-Celeb also held the first CN-Celeb Speaker Recognition Challenge (CNSRC) \footnote{\url{http://www.cnceleb.org/competition}} in 2022. 
% The data statistics of these both in-the-wild datasets are shown in Table \ref{}. \textcolor{red}{blabla}

This year, our team participated in the speaker recognition closed track in both the CNSRC and VoxSRC where only specified data can be used for training in the closed track.
% In this fixed track, only fixed training data from Voxceleb or CN-Celeb can be used. 
During these challenges, we adopted very similar strategies and achieved the 1$^{st}$ place in CNSRC 2022 challenge and 3$^{rd}$ place in VoxSRC 2022 challenge. 
% After comparing the top systems of these challenges in recent years \cite{thienpondt2020idlab,zhao2021speakin,chen2022sjtu}, we find that they have a lot in common. 
After comparing the top systems of these challenges in recent years \cite{thienpondt2020idlab,zhao2021speakin,chen2022sjtu}, we find that they all have a lot in common. However, the system descriptions of these top systems are usually not detailed enough and lack further analysis of the technologies they used.
% they only show the technologies they used but has very little analysis.

In this paper, we will first outline our challenge systems and show how to build a strong speaker verification system that can be used in the challenge. 
Then, from the perspectives of data augmentation, model backbone, loss functions and back-end scoring, we will give a further analysis of the technologies used in each module to verify their necessity and effectiveness. 

\section{System Pipeline}
\begin{figure*}[ht!]
  \centering
  \includegraphics[width=0.99\textwidth]{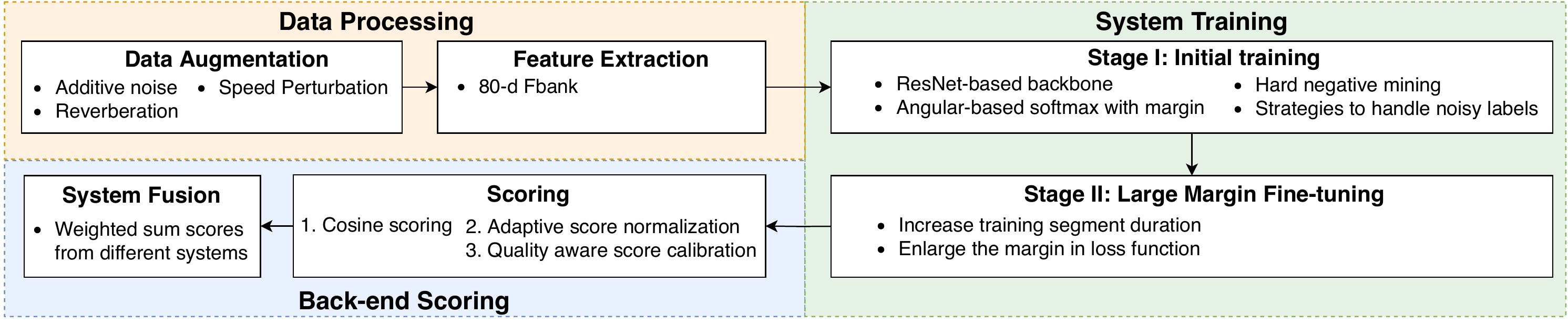}
  \caption{\textbf{The outline of our challenge system.} }
  \label{fig:whole_system}
  % \vspace{-0.5cm}
\end{figure*}
The outline of our system is shown in Figure \ref{fig:whole_system}. The system can be roughly divided into three parts: data processing, system training, and back-end scoring. Each part in the figure has a great influence on the final system performance. In this section, we will introduce a detailed description of each part.

\subsection{Data Processing}
% \subsubsection{Data Augmentation and Feature Extraction}
% In the challenge, we perform speed perturbation \cite{yamamoto2019speaker,wang2020dku,zhao2021speakin}, additive noise, and reverberation augmentation in an online manner. The detailed online augmentation pipeline is as follows:
Speed perturbation, additive noise, and reverberation augmentation are the most popular data augmentation methods in the challenge \cite{yamamoto2019speaker,wang2020dku,zhao2021speakin}, and we perform them in an online manner:
\begin{enumerate}
    \item Randomly sample a speed perturbation ratio from $\{1.0, 1.1, 0.9\}$ and do speed perturbation augmentation (ratio 1.0 means no speed perturbation). The speed perturbation will change the pitch of the audio and we consider the processed audio from a new speaker.
    % \item Decide whether to do noise augmentation with probability 0.6. If doing noise augmentation, randomly sampled a noise type from $\{\text{babble}, \text{noise}, \text{music}, \text{reverberation}\}$. The babble, noise and music audios are sampled from MUSAN \cite{musan2015} dataset. The reverberation impulse response is sampled from RIR\footnote{\url{https://www.openslr.org/28}} dataset.
    \item Decide whether to do noise augmentation with probability 0.6. If doing noise augmentation, randomly sample a noise type from $\{\text{babble}, \text{noise}, \text{music}, \text{reverberation}\}$, which are from MUSAN~\cite{musan2015} and RIR\footnote{\url{https://www.openslr.org/28}} dataset.
\end{enumerate}

% \subsubsection{Feature Extraction}
For feature extraction, we use torchaudio \cite{yang2022torchaudio} toolkit to extract the 80-dimensional Fbank and then do mean-normalization on it.
% In our challenge, we extract the 80-dimensional Fbank feature using torchaudio \cite{yang2022torchaudio} toolkit and do mean-normalization on it.

\subsection{System training}

\subsubsection{Embedding Extractor Backbone}
In the challenge, we mainly focus on ResNet-based r-vector as the embedding extractor which is implemented in \cite{zeinali2019but}. 
In addition, we also tune the channel or depth of ResNet to make it wider or deeper including ResNet101, ResNet152, ResNet221 and ResNet293~\cite{chen2022sjtu} for further performance improvements. 
The pooling function plays a vital role in embedding extractor~\cite{wang2021revisiting}. We also adopt the multi-query multi-head attention (MQMHA) pooling \cite{zhao2022multi} in part of our systems.

\subsubsection{Loss Function}
In our experiment, Additive Angular Margin (AAM) loss \cite{deng2019arcface,xiang2019margin} is used as the primary training objective to maximize the between-class distance and minimize the within-class distance. Besides, we also involve the Additive Margin (AM) loss \cite{wang2018additive,xiang2019margin} in our experiment for comparison. 
Besides, to alleviate noisy labels influence in the training set, we combine the Sub-center \cite{deng2020sub} with the AAM for robust training. The evaluation trial focuses more on the hard verification pairs for the speaker recognition challenge. Here, we also involve the Inter-TopK loss \cite{zhao2022multi} to add an extra penalty for $k$ closest centers to the example $x_i$. 
Besides, we also propose another hard example mining (HEM) loss which is defined as:
\begin{equation}
    L_{HEM}=-\log \frac{e^{s \cos \left(\theta_{i,y_i}+m\right)}}{e^{s \cos \left(\theta_{i,y_i}+m\right)}+\sum_{j=1, j \neq y_i}^N e^{s \phi(\theta_{i,j})}}
\end{equation}
where $\theta_{i,y_i}$ is the angle between the embedding and target classification center, and $\theta_{i,j}$ is the angle between the embedding and non-target classification center. To do hard sample mining, we detect the sample which has high similarity with the non-target classification center and adds an extra margin $m^{\prime}$ to $\theta_{i,j}$:
\begin{equation}
\phi\left(\theta_{i,j}\right) = 
\begin{cases}\cos (\theta_{i,j}-m^{\prime}) & \text {If } \cos \theta_{i,j} >  \cos \left(\theta_{i,y_i}+m\right)
\\ \cos \theta_{i,j} & \text {Otherwise. }\end{cases}
\end{equation}
$m^{\prime}$ here is set to 0.1 in our experiment.
\subsubsection{Training Strategy}
The training process in our challenge system is divided into two stages.

\noindent \textbf{Stage I: Initial Training}: In this stage, we use 2 seconds training segment. We set the margin and scale in AAM (or AM) to 0.2 and 32.0 respectively, and set the margin and $k$ in Inter-TopK loss to 0.06 and 5 respectively. For sub-center loss, we use 3 sub-centers for each class. In this stage, we iterate the training set for 150 epochs with SGD as the optimizer.
% with initial learning rate 0.1 and final learning rate 0.00005. 
During the training process, the learning rate is exponentially decreased from the initial 0.1 to the final 1e-5. 
% Besides, SGD is used as the optimizer in our experiment.

\noindent \textbf{Stage II: Large Margin Fine-tuning}: Large Margin fine-tuning (LM-FT) is first proposed in \cite{thienpondt2021idlab}, which aims to minimize the duration mismatch between the training and evaluation stage, and further optimize the with-class and between-class distance by enlarging the margin of loss function. In our experiment, we set the training segment duration to 6s in this stage and increase the margin to 0.5. The speed perturbation augmentation is abandoned here. Like the setup in \cite{zhao2021speakin}, we disable the Inter-TopK loss in this stage to make the training more stable. The learning rate here is initialized as 1e-4 and decreases to 2.5e-5 finally. Based on the pre-trained model from Stage I, we fine-tune it for only five epochs because the longer training segments make the model easy to overfit in this stage.
% Besides, we find it easy to overfit in the LM-FT stage because of the longer training segment. Based on the pre-trained model from last stage, we fine-tune the model for only 5 epoches with initial learning rate 0.0001 and final learning rate 0.000025.

\subsection{Back-end Scoring}
Because the angular-based softmax directly optimizes the speaker embedding in a hyper-sphere space, we use the cosine similarity to score the trials. Then, adaptive score normalization (as-norm) \cite{cumani2011comparison} is used to normalize the trial score. We average the embeddings from the same speaker in training set to construct the imposter cohort and set the imposter cohort size to 600.   Finally, we use the quality-aware score calibration \cite{thienpondt2021idlab} to introduce some extra information to calibrate the score. The Quality Measuring Function (QMF) attributes used in our experiments include:
\begin{itemize}
    \item The magnitude of enroll and test embeddings.
    \item The duration of enroll and test utterances.
    \item The mean value of imposter cohort for enroll and test utterances.
\end{itemize}

\section{Experimental Setup}
\subsection{VoxSRC 2022}
The Voxceleb2 dev \cite{chung2018voxceleb2} set is used as the training set, which contains 5994 speakers. The evaluation trials in our experiment include three cleaned version trials Vox1-O, Vox1-E and Vox1-H constructed from 1251 speakers in Voxceleb1 \cite{nagrani2017voxceleb}, and the validation trials from VoxSRC 2021 and VoxSRC 2022. Besides, we construct a trial with 30k pairs from the Voxceleb2 dev set following the strategy proposed in \cite{thienpondt2021idlab} to train the score calibration model. In the experimental analysis part, we use ResNet34 with statistic pooling, AAM loss function and cosine similarity scoring plus as-norm and score calibration as the default setup. We don't involve the large margin fine-tuning strategy without specific notation.

\subsection{CNSRC 2022}
The dev set of CN-Celeb1 \cite{fan2020cn} and CN-Celeb2 \cite{li2022cn} are used as the training set, which contains 2787 speakers in total. Because there are many short utterances less than 2s in CN-Celeb dataset~\cite{li2022cn}, we first concatenate the short utterances from the same genre and same speaker to make them longer than 5s. It should be noted that we only do this operation on the training set. Then, the training utterance number is reduced from 632,740 to 508,228. We report results on CN-Celeb evaluation trial which is also used as the challenge evaluation set.
% The official CN-Celeb evaluation is also used as the challenge evaluation set and we also report results on it. 
% It should be noted that we didn't apply score calibration for CNSRC 2022. We find the score calibration can benefit the EER but degrade the minDCF (p=0.01). Because the minDCF is the main evaluation metric of the challenge, we abandon the score calibration in this challenge.
It should be noted that we abandon the score calibration here because we find that the score calibration can benefit the EER but degrade the minDCF which is the main evaluation metric of this challenge.
Besides, there are multiple utterances for each enrollment speaker in CNSRC 2022. We average all the embeddings belonging to each enrollment speaker to get the final speaker embedding.

\section{Results and Analysis}
\subsection{Data Augmentation Methods Comparison}
\begin{table}[ht!]
% \vspace{-10pt}
% \footnotesize
\centering
\caption{\textbf{Voxceleb EER (\%) results comparison between different augmentation methods.} N+R: additive noise and reverberation. Perturb (S): audio perturbation by changing audio speed. Perturb (P): audio perturbation by changing audio pitch. Perturb (S+P): audio perturbation by changing both audio speed and pitch.}
\begin{adjustbox}{width=.42\textwidth,center}
\begin{threeparttable}
\begin{tabular}{lcccc}
    \toprule 
  
      Aug Type &  Vox1-O & Vox1-E &  Vox1-H \\
      \hline
    %   SpecAug & & \\
      N + R & 0.947 & 1.098 & 2.002 \\
      N + R + SpecAugment & 1.064 & 1.091 & 2.004 \\
      N + R + Perturb (S) & 0.973 & 1.113 & 2.033 \\
      N + R + Perturb (P) & 0.867 & \textbf{0.984} & 1.781 \\
      N + R + Perturb (S + P) & 0.861 & 0.996 & \textbf{1.767}\\
      N + R + Perturb (S + P) $^*$ & \textbf{0.803} & 1.001 & 1.777 \\
    \bottomrule
\end{tabular}
\begin{tablenotes}\footnotesize
\item $^*$: applying speed perturbation with ratio $\{0.8, 0.9, 1.0, 1.1, 1.2\}$
% \item $^\dagger$: applying extra pitch perturbation $P$ in the fine-tuning process
\end{tablenotes}
\end{threeparttable}
\label{table:res_diff_aug}
\end{adjustbox}
\end{table}

In this section, we first analyze the effect of different data augmentation methods. The effectiveness of the additive noise (N) and reverberation (R) have been extensively verified \cite{snyder2017deep,snyder2018x} and we consider it as the baseline in this experiment. The results are shown in Table \ref{table:res_diff_aug}. Apart from the additive noise and reverberation, we also investigate the effectiveness of SpecAugment \cite{park2019specaugment,wang2020investigation} and speed perturbation. It should be noted that the speed perturbation used in many challenge systems \cite{wang2020dku,zhao2021speakin} is just one case of audio perturbation method \cite{yamamoto2019speaker}. They implement it using `sox speed' \footnote{\url{http://sox.sourceforge.net/}} command, which will change the speed and pitch (S+P) of audio. However, there is no detailed analysis of whether speed augmentation or pitch augmentation is more useful. Here, we also apply the audio perturbation by only changing audio speed (S) using `sox tempo' command and the audio perturbation by only changing audio pitch (P) using `sox speed' and `sox tempo' command sequentially \footnote{We can also use `sox pitch' command to only change the audio pitch.}.

Although some previous works \cite{wang2020investigation} have shown that it is useful to use SpecAugment alone, the results in Table \ref{table:res_diff_aug} show that combining SpecAugment with additive noise and reverberation cannot obtain further benefits. 
Besides, the audio perturbation by only changing the audio speed cannot improve the system performance, which demonstrates the gain of speed perturbation (S+P) comes from the audio pitch augmentation. In many challenge systems, they only speed up or slow down the audio speed with ratios of 1.1 and 0.9 respectively. Here, we additionally explore the speed ratios 1.2 and 0.8 to generate more diverse training data. However, we find that the performance from speed perturbation has been saturating and the additional speed perturbation ratio cannot bring more performance improvement. In the following experiments, we will use `` N + R + Perturb (S + P)'' in Table \ref{table:res_diff_aug} as the default setup.

\subsection{Acoustic Feature Comparison}

In this section, we analyze the performance of different acoustic features and list the results in Table \ref{table:res_diff_feature}. 
% The MFCC feature is the most widely used in the speaker field. However, in the speaker recognition challenge, more researchers turned their attention to Fbank features and used higher dimensional Fbank. The results show that all the Fbank features perform better than the MFCC feature. Such a phenomenon is reasonable because the MFCC feature is transformed from the Fbank feature, and some information may be lost during this transformation process. Besides, the higher dimensional Fbank feature seems to perform better than the lower dimensional one. Finally, we add additional pitch information to the Fbank feature and there is no further improvement. We will use the 80-dimension Fbank feature in the following experiments.
The MFCC feature is actually the most widely used feature in the speech field. However, for speaker recognition challenge, more researchers turned their attention to Fbank features because the MFCC feature is transformed from the Fbank feature and some information may be lost during this transformation process. In Table \ref{table:res_diff_feature}, all the Fbank features perform better than the MFCC feature, demonstrating this point. Besides, the higher dimensional Fbank feature seems to perform better than the lower dimensional one when we compare the Fbank 40d and 80d. And if we continue to increase to 96d, although the performance will be slightly improved, it will also bring more computation. Finally, we also try to add additional pitch information with Fbank feature but there is no further gain. Considering the trade-off between performance and computation, we will use 80d Fbank in the following experiments.

\begin{table}[t!]
% \footnotesize
\centering
\caption{\textbf{Voxceleb EER (\%) results comparison between different acoustic features.}}
\begin{adjustbox}{width=.40\textwidth,center}
\begin{threeparttable}
\begin{tabular}{lcccc}
    \toprule 
      Acoustic Feature &  Vox1-O & Vox1-E &  Vox1-H \\
      \hline
      MFCC-80d & 1.292 & 1.351 & 2.436 \\
      Fbank-40d & 0.941 & 1.147 & 2.104 \\
      Fbank-80d & \textbf{0.861} & 0.996 & 1.767 \\
      Fbank-96d & 0.931 & \textbf{0.953} & \textbf{1.741} \\
      Fbank-80d + Pitch & 0.862 & 0.984 & 1.783  \\
    \bottomrule
\end{tabular}
\end{threeparttable}
\label{table:res_diff_feature}
\end{adjustbox}
\end{table}

\subsection{Scoring Method}
\begin{table}[ht!]
% \vspace{-10pt}
% \footnotesize
\centering
\caption{\textbf{Voxceleb EER (\%) results comparison between different scoring methods.} PLDA is trained on Voxceleb2 dev set.}
\begin{adjustbox}{width=.40\textwidth,center}
\begin{threeparttable}
\begin{tabular}{lcccc}
    \toprule 
  
      Scoring Method &  Vox1-O & Vox1-E &  Vox1-H \\
      \hline
      PLDA & 1.633 & 1.723 & 2.857 \\
      Cosine & 1.058 & 1.147 & 2.087\\
       + AS-Norm & 0.920 & 1.048 & 1.874 \\
       ++ Score Calibration & \textbf{0.861} & \textbf{0.996} & \textbf{1.767} \\
       
    \bottomrule
\end{tabular}

\end{threeparttable}
\label{table:res_diff_scoring}
\end{adjustbox}
\end{table}

In this section, we analyze the effect of different scoring methods and results are shown in Table \ref{table:res_diff_scoring}. PLDA used to be a popular scoring method in speaker recognition field. However, with the advent of angular-based softmax \cite{deng2019arcface,wang2018additive} which can optimize the speaker embedding in a hyper-sphere space, cosine scoring methods begin to dominate. The results show that the performance of PLDA scoring has lagged far behind cosine scoring. 
% Of course, we have noticed that some recent works have proposed new strategies to improve PLDA's performance \cite{peng2022unifying,wang2022scoring}. Because the new PLDA methods have no obvious improvement over the cosine scoring, we still mainly use the cosine scoring in the challenge. 
After cosine scoring, we apply as-norm and quality-aware score calibration and these two compensation strategies can make consistent improvements. We will also use the cosine scoring + as-norm + score calibration strategy in the following experiments.

\subsection{Loss Function Comparison and Training Strategy}

\begin{table*}[ht!]
% \renewcommand\thetable{5}
% \footnotesize
\centering
\caption{\textbf{Voxceleb and VoxSRC results comparison between different embedding extractor backbones.} `-64' denotes the ResNet with base channel number 64 and our default setup is 32. Large margin fine-tuning is applied for all the systems. The ResNet models with statistic pooling are trained with AAM loss and the ResNet models with MQMHA pooling are trained with AAM + Sub-center + Inter-Topk loss.}
\begin{adjustbox}{width=.98\textwidth,center}
\begin{threeparttable}
\begin{tabular}{llcccccccccccc}
    \toprule 
      \multirow{2}{*}{Index} & \multirow{2}{*}{Model Type} & \multirow{2}{*}{Pooling}  & \multirow{2}{*}{Param \#} & \multicolumn{2}{c}{Vox1-O} & \multicolumn{2}{c}{Vox1-E} & \multicolumn{2}{c}{Vox1-H} & \multicolumn{2}{c}{VoxSRC21-val} & \multicolumn{2}{c}{VoxSRC22-val} \\
      \cline{5-14}
      & & & & DCF$_{0.01}$ & EER & DCF$_{0.01}$ & EER  & DCF$_{0.01}$ & EER & DCF$_{0.05}$ & EER  & DCF$_{0.05}$ & EER \\
      \hline
      S1 & ECAPA (c=512) & ASP & 6.20M    & 0.0901 & 0.771 & 0.1051 & 0.962 & 0.1670 & 1.784 & 0.1911 & 3.540 & 0.1676 & 2.499 \\
      S2 & ECAPA (c=1024) & ASP & 14.7M  & 0.0802 & 0.606 & 0.0889 & 0.813 & 0.1556 & 1.594 & 0.1841 & 3.267 & 0.1641 & 2.398 \\
      S3 & ResNet34 & Statistic & 6.63M  & 0.0635 & 0.654 & 0.0920 & 0.824 & 0.1424 & 1.456 & 0.1434 & 2.574 & 0.1456 & 2.085 \\
      S4 & ResNet101 & Statistic & 15.9M & 0.0404 & 0.505 & 0.0666 & 0.651 & 0.1101 & 1.163 & 0.1162 & 2.025 & 0.1154 &  1.691  \\
      S5 & ResNet152 & Statistic & 19.8M & 0.0341 & 0.415 & 0.0623 & 0.606 & 0.1027 & 1.101 & 0.1189 & 2.052 & 0.1077 &  1.569  \\
      S6 & ResNet34 & MQMHA  &  8.61M    & 0.0471 & 0.675 & 0.0843 & 0.771 & 0.1353 & 1.369 & 0.1417 & 2.411 & 0.1408 &  1.955  \\
      S7 & ResNet34-c64 & MQMHA & 27.8M  & 0.0510 & 0.638 & 0.0771 & 0.756 & 0.1275 & 1.353 & 0.1429 & 2.604 & 0.1375 & 2.001 \\
      S8 & ResNet101 & MQMHA & 23.8M     & 0.0442 & 0.425 & 0.0615 & 0.602 & 0.1003 & 1.051 & 0.1050 & 1.885 & 0.1029 & 1.551 \\
      S9 & ResNet101-c64 & MQMHA & 68.7M & 0.0335 & 0.388 & 0.0575 & 0.576 & 0.0964 & 1.044 & 0.1124 & 1.961 & 0.1034 & 1.566 \\
      S10 & ResNet152 & MQMHA & 27.7M     & \textbf{0.0321} & 0.378 & 0.0549 & 0.552 & 0.0898 & 0.980 & \textbf{0.0967} & \textbf{1.765} & 0.1036 & 1.457 \\
      S11 & ResNet221 & MQMHA & 31.6M     & 0.0357 & \textbf{0.330} & \textbf{0.0539} & \textbf{0.535} & \textbf{0.0855} & \textbf{0.966} & 0.1009 & 1.795 & \textbf{0.0976} & \textbf{1.420} \\
      \hline
      S1-S11 & Fusion & - & -     & 0.0282 & 0.303 & 0.0483 & 0.491 & 0.0795 & 0.892 & 0.0931 & 1.645 & 0.0898 & 1.330 \\
    \bottomrule
\end{tabular}
\end{threeparttable}
\label{table:res_diff_backbone}
\end{adjustbox}
\vspace{-10pt}
% \vspace{-0.5cm}
\end{table*}

This section compares the results of different loss functions and training strategies. The results are shown in Table \ref{table:res_loss}. 
% First, we find that AAM performs slightly better the AM loss. 
First, we compare the AM and AAM loss and find that they both achieve comparable results. Then we choose AAM as our default loss.
% For exploring hard trials, we add an extra penalty to the hard negatives using HEM and Inter-TopK loss. 
The Sub-center strategy is adopted here to alleviate the noisy labels in training data and obtains performance gain compared with baseline AAM.
To mine the hard samples during training, our proposed HEM loss and Inter-Topk are both investigated here. 
The results show that the HEM loss and Inter-Topk can both improve the performance significantly, especially on the challenge's harder trial. In addition, our proposed HEM achieves the best EER on VoxSRC2021-val trial. 
% Based on the Inter-TopK loss, we further add the Sub-center module and get the consistent improvement. 
Finally, we apply the large margin fine-tuning (LM-FT) on the pre-trained model. It can be seen that the performance improvement brought by the LM-FT is huge. We will also apply the LM-FT in the following experiments.

\begin{table}[ht!]
% \vspace{-5pt}
% \renewcommand\thetable{4}
% \footnotesize
\centering
\caption{\textbf{Voxceleb and VoxSRC EER (\%) results comparison between different loss functions.} The LM-FT is performed on the model with Inter-Topk loss.}
\begin{adjustbox}{width=.45\textwidth,center}
\begin{threeparttable}
\begin{tabular}{lcccc}
    \toprule 
      Loss Functions &  Vox1-O & Vox1-E &  Vox1-H & VoxSRC21-val\\
      \hline
      AM & 0.840 & 0.987 & 1.796 & 3.453\\
      AAM & 0.861 & 0.996 & 1.767 & 3.450 \\
    %   + HEM & 0.861 & 0.983 & 1.766 & 3.207 \\ 
    %   + Inter-TopK & 0.824 & 0.968 & 1.724 & 3.137 \\ 
      + Sub-center & 0.824 & 0.985 & 1.733 & 3.340\\
      ++ HEM & \textbf{0.782} & 0.970 & 1.684 & \textbf{3.060} \\
      ++ Inter-TopK & \textbf{0.782} & \textbf{0.936} & \textbf{1.658} & 3.153 \\ 
      \hline
      LM-FT & 0.649 & 0.797 & 1.394 & 2.429 \\
    \bottomrule
\end{tabular}
\end{threeparttable}
\label{table:res_loss}
\end{adjustbox}
% \vspace{-5pt}
\end{table}

\subsection{Embedding Extractor Backbone Comparison}
% Here, we compare the different speaker embedding extractor backbones and list the results in Table \ref{table:res_diff_backbone}. ResNet-based embedding extractor and TDNN-based embedding extractor are two kinds of widely used backbones, and we mainly focus on them or their variants. 
Except for the ResNet-based backbone, we also involve a very popular TDNN-based model, ECAPA-TDNN~\cite{desplanques2020ecapa}, for comparison. Besides, we also apply different pooling methods for the ResNet-based model. 
% For TDNN-based backbone, we use the ECAPA-TDNN, which is the most powerful TDNN model. For ResNet-based model, we tried many configurations with different depths and widths (channel number). We also apply different pooling methods for ResNet-based model. 
From the results in Table \ref{table:res_diff_backbone}, we find that the ResNet-based systems perform much better than the ECAPA-TDNN systems. Although the ResNet34 has similar performance with ECAPA-TDNN (c=1024) following the configuration in the original ECAPA-TDNN paper \cite{desplanques2020ecapa}\footnote{In our experiment, ResNet34 has 2.048\% EER on Vox1-H and ECAPA-TDNN (c=1024) has 2.017\% EER on Vox1-H following the setup in \cite{desplanques2020ecapa}.}, the ResNet-based model shows higher performance upper bound after adding more extensive data augmentation and advanced training strategies. Besides, we find that making ResNet deeper can improve the performance a lot but making ResNet wider has little effect on the performance. Further, we add the MQMHA pooling and Sub-center+Inter-TopK loss functions to the ResNet training and achieve further performance improvement. Finally, we fuse all the systems' scores in Table \ref{table:res_diff_backbone} based on the performance on trial VoxSRC22-val to get the fusion system.

\subsection{Results for CNSRC}
\begin{table}[ht!]
\vspace{-5pt}
% \footnotesize
\centering
\caption{\textbf{Results on CNSRC 2022.} Without specific notation, the models using statistic pooling and are trained with AAM loss.}
\begin{adjustbox}{width=.43\textwidth,center}
\begin{threeparttable}
\begin{tabular}{llcccc}
    \toprule 
      Index & Model & Param \# &  DCF$_{0.01}$ & EER (\%) \\
      \hline
      S1 & ResNet34 $^*$         & 6.63M & 0.3958 & 7.981 \\
      S2 & ResNet34              & 6.63M & 0.3707 & 6.590 \\
      S3 & ResNet34  $^\dagger$     & 6.63M & 0.3593 & 6.702  \\
      S4 & ResNet152             & 19.8M & 0.3386 & 5.762 \\
      S5 & ResNet221             & 23.8M & 0.3270 & \textbf{5.543} \\
      S6 & ResNet293             & 28.6M & \textbf{0.3202} & 5.553  \\
      S7 & DF-ResNet      & 14.8M & 0.3361 & 6.279 \\
      \hline
      S8 & ResNet34 + LM-FT $^\dagger$  & 6.63M & 0.3458 & 6.319   \\
      S9 & ResNet34 + LM-FT         & 6.63M & 0.3543 & 6.221 \\
      S10 & ResNet152 + LM-FT        & 19.8M & 0.3251 & 5.452 \\
      S11 & ResNet221 + LM-FT        & 23.8M & 0.3179 & 5.284 \\
      S12 & ResNet293 + LM-FT        & 28.6M & \textbf{0.3164} & \textbf{5.227} \\
      S13 & DF-ResNet + LM-FT & 14.8M & 0.3185 & 6.117  \\
      \hline
      S9-S13& Fusion  & - &  0.2975 & 4.911  \\
      - & SpeakerIn System \cite{zheng2022speakin} & - &  0.3185 & 5.953  \\
      -  & STAP System \cite{shiwei}  & - &  0.3399 & 5.728  \\
       
    \bottomrule
\end{tabular}
\begin{tablenotes}\footnotesize
\item $^*$: did not apply speed perturbation.
\item $^\dagger$: using MHMQA pooling and AAM + Sub-center + Inter-TopK loss.
\end{tablenotes}
\end{threeparttable}
\label{table:res_cnceleb}
\end{adjustbox}
\vspace{-5pt}
\end{table}

In the previous sections, we have introduced our systems in VoxSRC 2022, which have very competitive performance. Here, similar systems are applied in the CNSRC 2022 and the results are given in Table \ref{table:res_cnceleb}. Besides, we also involved the DF-ResNet \cite{liu22h_interspeech} in this challenge. From the results, we find that the speed perturbation augmentation, deep ResNet and large margin fine-tuning all play an important role. Finally, we weighted sum the scores of all the LM-FT systems (except the system denoted with $^\dagger$) based on their performance to get the fusion system.
Our team also achieved the 1$^{st}$ place in the CNSRC 2022, which further confirms that our system is not only strong but also generalizable to different scenarios.

\section{Conclusion}
In most system reports, the authors usually only provide a description of their systems but lack an effective analysis of their methods. 
In this paper, based on our winner systems in CNSRC 2022 and VoxSRC 2022, we outline how to build a strong speaker verification challenge system. Then, from the perspectives of data augmentation, model backbone, loss functions, and back-end scoring, we share some experiences learned from these challenges and give a further analysis of the technologies used to verify their necessity and effectiveness.

\bibliographystyle{IEEEtran}
\bibliography{refs}

\end{document}